\documentclass[useAMS,usenatbib]{mn2e}

\usepackage{latexsym,graphicx,natbib}

\newcommand\kms{{\rm\,km\,s^{-1}}}

\newcommand\msun{\rm\,M_\odot}
\newcommand\rsun{\rm\,R_\odot}

\def\apgt{\ {\raise-.5ex\hbox{$\buildrel>\over\sim$}}\ }
\def\aplt{\ {\raise-.5ex\hbox{$\buildrel<\over\sim$}}\ }

\title[HD\,271791: dynamical ejection scenario]{HD\,271791:
dynamical versus binary-supernova ejection scenario}
\author[V.\,V.\,Gvaramadze]
       {V.\,V.\,Gvaramadze\thanks{E-mail: vgvaram@mx.iki.rssi.ru}\\
       Sternberg Astronomical Institute, Moscow State University,
       Universitetskij Pr. 13, Moscow 119992, Russia}
\begin{document}

\date{Accepted 2009 February 26. Received 2009 February 21; in original form 2009 January 16}

\maketitle

\begin{abstract}
The atmosphere of the extremely high-velocity ($530-920 \, \kms$)
early B-type star HD\,271791 is enriched in $\alpha$-process
elements, which suggests that this star is a former secondary
component of a massive tight binary system and that its surface was
polluted by the nucleosynthetic products after the primary star
exploded in a supernova. It was proposed that the (asymmetric)
supernova explosion unbind the system and that the secondary star
(HD\,271791) was released at its orbital velocity in the direction
of Galactic rotation. In this Letter, we show that to explain the
Galactic rest-frame velocity of HD\,271791 within the framework of
the binary-supernova scenario, the stellar remnant of the supernova
explosion (a $\la 10 \, \msun$ black hole) should receive an
unrealistically large kick velocity of $\geq 750-1200 \, \kms$. We
therefore consider the binary-supernova scenario as highly unlikely
and instead propose that HD\,271791 attained its peculiar velocity
in the course of a strong dynamical three- or four-body encounter in
the dense core of the parent star cluster. Our proposal implies that
by the moment of encounter HD\,271791 was a member of a massive
post-supernova binary.
\end{abstract}


\section{Introduction}
%
HD\,271791 (also MO\,88) is a B2III star (Carozzi 1974) located at a
high Galactic latitude of $\simeq 30\degr$. The large separation of
HD\,271791 from the Galactic plane and its very high heliocentric
radial velocity of $\ga 400 \, \kms$ (Carozzi 1974; Kilkenny \&
Muller 1989) suggest that this star is an extremely high-velocity
runaway star. Subsequent proper motion measurements showed that
HD\,271791 originated on the periphery of the Galactic disc (at a
galactocentric distance of $\ga 15$ kpc) and that its Galactic
rest-frame velocity is $\simeq 530-920 \, \kms$ (Heber et al. 2008).
Velocities of this order of magnitude are typical of the so-called
hypervelocity stars (HVSs) -- the ordinary stars moving with
peculiar velocities exceeding the escape velocity of our Galaxy
(Brown et al. 2005; Edelmann et al. 2005; Hirsch et al. 2005). The
existence of the HVSs was predicted by Hills (1988), who showed that
close encounter between a tight binary system and the supermassive
black hole (BH) in the Galactic Centre could be responsible for
ejection of one of the binary components with a velocity of several
$1000 \, \kms$ (see also Yu \& Tremaine 2003). It is therefore
plausible that some HVSs were produced in that way (Gualandris,
Portegies Zwart \& Sipior 2005; Baumgardt, Gualandris \& Portegies
Zwart 2006; Levin 2006; Sesana, Haardt \& Madau 2006; Ginsburg \&
Loeb 2006; Lu, Yu \& Lin 2007; L\"{o}ckmann \& Baumgardt 2008).
Interestingly, HD\,271791 is the only HVS whose birth place was
constrained by direct proper motion measurements and the origin of
just this star cannot be associated with the Galactic Centre.

An alternative explanation of the origin of HVSs and other
high-velocity objects (e.g. the hyperfast neutron stars; see
Chatterjee et al. 2005; Hui \& Becker 2006) is that they (or their
progenitors) attained peculiar velocities in the course of strong
dynamical three- or four-body encounters in young and dense star
clusters located in the Galactic disc (Gvaramadze 2006a, 2007;
Gvaramadze, Gualandris \& Portegies Zwart 2008) or in the Large
Magellanic Cloud (Gualandris \& Portegies Zwart 2007).

Przybilla et al. (2008) found that the atmosphere of HD\,271791 is
enriched in several $\alpha$-process elements. To explain this
enrichment, they suggested that HD\,271791 was a secondary of a
massive tight binary and that its surface was polluted by the
nucleosynthetic products after the primary star exploded in a
supernova (SN). They also suggested that the binary was disrupted by
(asymmetric) SN explosion and assumed that HD\,271791 was released
at its orbital velocity. In this Letter, we show that, to explain
the peculiar velocity of HD\,271791 within the framework of the
binary-SN scenario, the stellar remnant of the SN explosion (a $\la
10 \, \msun$ BH, according to Przybilla et al. 2008) should receive
an unrealistically large kick velocity of $\geq 750-1200 \, \kms$.
We therefore consider the binary-SN scenario as highly unlikely and
instead propose that HD\,271791 attained its extremely high peculiar
velocity in the course of a strong dynamical three- or four-body
encounter in the dense core of the parent star cluster. Our proposal
implies that by the moment of encounter HD\,271791 was a member of a
massive post-SN binary.

\section{HD\,271791 as a former secondary component of a massive tight binary}

The spectral analysis of HD\,271791 by Przybilla et al. (2008)
revealed that the Fe abundance in its atmosphere is subsolar and
that the $\alpha$-process elements are enhanced. The first finding
is consistent with the origin of HD\,271791 in the metal-poor
outskirts of the Galactic disc, while the second one suggests that
this star was a secondary component of a massive tight binary (see
above). Przybilla et al. (2008) believe that the binary-SN explosion
could be responsible not only for the $\alpha$-enhancement in
HD\,271791 but also for the extremely high space velocity of this
star. Below, we outline their scenario.

The large separation of HD\,271791 from the Galactic plane ($\simeq
10$ kpc) along with the proper motion measurements (Heber et al.
2008) implies that the time-of-flight of this $11\pm 1 \, \msun$
star is comparable to its evolutionary lifetime of $25\pm 5$ Myr,
which in turn implies that the star was ejected within several Myr
after its birth in the Galactic disc. The ejection event was
connected with disruption of a massive tight binary following the SN
explosion. The original binary was composed of a primary star of
mass of $\ga 55 \, \msun$ and an early B-type secondary
(HD\,271791), so that the SN explosion and the binary disruption
occurred early in the lifetime of HD\,271791. The system was close
enough to go through the common-envelope phase before the primary
exploded in a SN. During the common-envelope phase, the primary star
lost most of its hydrogen envelope and the binary became a tight
system composing of a Wolf-Rayet star and an early B-type
main-sequence star\footnote{For an alternative channel for the
formation of very tight massive binaries see de Mink et al. (2009)}.
At the moment of SN explosion, the mass of the primary star was $\la
20 \, \msun$ and the binary semimajor axis was $\sim 14 \rsun$ (that
corresponds to the orbital velocity of the secondary of $\la 420 \,
\kms$). The exploded star expelled $\simeq 10 \, \msun$ of its mass
while the remaining mass collapsed to a $\la 10 \, \msun$ BH. The SN
explosion was asymmetric enough to disrupt the system. Przybilla et
al. (2008) assumed that HD\,271791 was released at its orbital
velocity and that at the time of binary disruption the vector of the
orbital velocity was directed by chance along the Galactic rotation
direction. The first assumption is based on the wide-spread
erroneous belief that runaways produced from a SN in a binary system
have peculiar velocities comparable to their pre-SN orbital
velocities. The second assumption is required to explain the
difference between the assumed space velocity from the binary
disruption and the Galactic rest-frame velocity of HD\,271791
(provided that the latter is on the low end of the observed range
$530-920 \, \kms$).

In the next Section, we discuss the conditions under which the
secondary star could be launched into free flight at a velocity
equal to its pre-SN orbital one.

\section{HD\,271791: binary-supernova scenario}

One of two basic mechanisms producing runaway stars is based on a SN
explosion in a massive tight binary system (Blaauw 1961). (The
second one is discussed in Section\,4.) After the primary star
exploded in a SN, the binary system could be disintegrated if the
system lost more than half of its pre-SN mass (Boersma 1961) and/or
the SN explosion was asymmetric (so that the stellar remnant, either
a neutron star or a BH, received at birth a kick velocity exceeding
the escape velocity from the system; Stone 1982; Tauris \& Takens
1998).

In the case of binary disruption following the symmetric SN
explosion, the stellar remnant is released at its orbital velocity,
while the space velocity of the secondary star, $V_{\rm sec}$, is
given by (Boersma 1961; Radhakrishnan \& Shukre 1985; Tauris \&
Takens 1998)
\begin{equation}
V_{\rm sec} = \sqrt{1-2{m_1 +m_2 \over m_1 ^2}} \, V_{\rm orb} \, ,
\end{equation}
where $m_1 > 2 + m_2 , m_1 =M_1 /M_{\rm co} , m_2 =M_2 /M_{\rm co}$,
$M_1$ and $M_2$ are the pre-SN masses of the primary and the
secondary stars, $M_{\rm co}$ is the mass of the compact object
formed in the SN explosion, $V_{\rm orb} = [GM_1 ^2 /(M_1
+M_2)a]^{1/2}$ is the orbital velocity of the secondary star, $G$ is
the gravitational constant and $a$ is the binary semimajor axis. It
follows from equation\,(1) that $V_{\rm sec} \simeq V_{\rm orb}
\simeq (GM_1 /a)^{1/2}$ if $M_1 >> M_2 , M_{\rm co}$.

The above consideration shows that the secondary star could achieve
a high peculiar speed if one adopts a large pre-SN mass of the
primary (i.e. $M_1 >> M_{\rm co}$). But, the stellar evolutionary
models suggest that the pre-SN masses of stars with initial
(zero-age main-sequence) masses, $M_{\rm ZAMS}$, from 12 to $120 \,
\msun$ do not exceed $\sim 10-17 \, \msun$ (Schaller et al. 1992;
Vanbeveren, De Loore \& Van Rensbergen 1998; Woosley, Heger \&
Weaver 2002; Meynet \& Maeder 2003). Using these figures and
assuming that the pre-SN binary is as tight as possible (i.e. the
secondary main-sequence star is close to filling its Roche lobe),
one can estimate the maximum possible velocity achieved by a runaway
star in the process of binary disruption following the symmetric SN
explosion. Assuming that the SN explosion left behind a neutron star
(i.e. $M_{\rm co} =1.4 \, \msun$) and adopting $M_1 =10-17 \,
\msun$, one has that a $3 \, \msun$ secondary star could attain a
peculiar velocity of $\simeq 300-500 \kms$, while a $10 \, \msun$
star could be ejected with a speed of $\simeq \, 350 \kms$.

Note that the pre-SN mass is maximum for stars with $M_{\rm ZAMS}
\simeq 20-25 \, \msun$ and $\ga 80 \, \msun$ [see Fig.\,6 of Meynet
\& Maeder (2003)]. In the first case, the SN explosion leave behind
a neutron star, while in the second one the stellar SN remnant is a
BH of mass $M_{\rm co} \geq 5 \, \msun$ (e.g. Woosley et al. 2002;
Eldridge \& Tout 2004). The large separation of HD\,271791 from the
Galactic plane implies that this massive star was ejected very soon
after its birth in the Galactic disc (see Section\,2). From this, it
follows that to explain the space velocity of HD\,271791 within the
framework of the binary-SN scenario one should assume that the
primary was a short-lived very massive star (Przybilla et al. 2008).
In this case, the stellar SN remnant is a BH and the SN ejecta is
not massive enough to cause the disruption of the binary system.

\begin{figure}
\includegraphics[width=8cm]{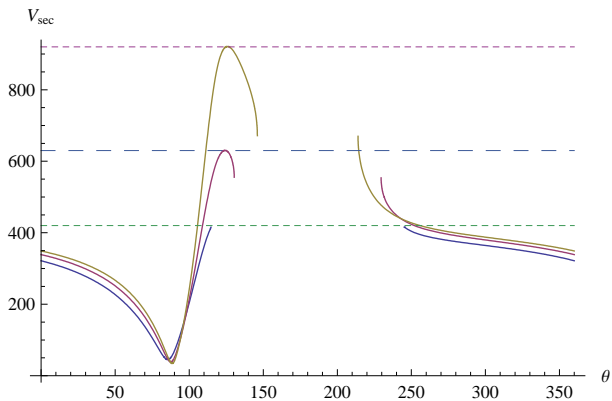}
\centering \caption{The peculiar velocity
of the secondary star (HD\,271791) as a function of the angle, $\theta$,
between the kick vector and the direction of motion of the exploding
star and the magnitude of the kick [$750 \, \kms$ (blue line),
$1000 \, \kms$ (red line), $1200 \, \kms$ (yellow line)]. The discontinuities in the
curves correspond to a range of angles $\theta$ for which the system remains
bound. The horizontal long-dashed line indicates the Galactic rest-frame
velocity of HD\,271791 of $630 \, \kms$, corresponding to the "best"
proper motion given in Heber et al. (2008). The horizontal short-dashed
lines indicate the orbital velocity of HD\,271791 of $420 \, \kms$
[suggested by the scenario of Przybilla et al. (2008)] and the maximum
possible Galactic rest-frame velocity of HD\,271791 of $920 \, \kms$ (Heber et al.
2008). See text for details.}
\end{figure}

According to Przybilla et al. (2008), the pre-SN mass of the primary
star was $\la 20 \, \msun$ (i.e. somewhat larger than the maximum
mass predicted by the stellar evolutionary models; see above) and
the SN explosion left behind a BH of mass $\la 10 \, \msun$
(comparable to the mass of the secondary star, HD\,271791), i.e. the
system lost less than a half of its mass. Thus, to disrupt the
binary, the SN explosion should be asymmetric. In this case, the
space velocities of the BH and HD\,271791 depend on the magnitude
and the direction of the kick imparted to the BH at birth (Tauris \&
Takens 1998). To estimate $V_{\rm sec}$, one can use
equations\,(44)-(47) and (54)-(56) given in Tauris \& Takens (1998).
It follows from these equations that $V_{\rm sec}$ is maximum if the
vector of the kick velocity does not strongly deviate from the
orbital plane of the binary and is directed nearly towards the
secondary, i.e. the angle, $\theta$, between the kick vector and the
direction of motion of the exploding star is $\sim \theta _{\ast} =
\arccos (-v/w)$, where $v=[G(M_1 +M_2 )/a]^{1/2}$ is the relative
orbital velocity and $w$ is the kick velocity (see Gvaramadze
2006b).

Fig.\,1 shows how the direction and the magnitude of the kick affect
$V_{\rm sec}$. The three solid lines represent $V_{\rm sec}$
calculated for the binary parameters suggested by Przybilla et al.
(2008) and three kick magnitudes of $750 \, \kms$ (blue line), $1000
\, \kms$ (red line) and $1200 \, \kms$ (yellow line). One can see
that to launch HD\,271791 at its pre-SN orbital velocity $V_{\rm
orb} \simeq 420 \, \kms$ the kick imparted to the BH should be at
least as large as $750 \, \kms$. In fact, the kick magnitude should
be much larger since for kicks $\simeq 750 \, \kms$ the kick
direction must be carefully tuned (see Fig\,1), i.e. $\theta$ should
be either $\simeq 115\degr$ or $\simeq 245\degr$ (note that for
$115\degr \la \theta \la 245\degr$ the binary system remains bound).
The even larger kicks of $\geq 1000$ and $\geq 1200 \, \kms$ are
required to explain the Galactic rest-frame velocities of HD\,271791
of $630$ and $920 \, \kms$ [corresponding, respectively, to the
"best" and the maximum proper motions given in Heber et al. (2008);
see also Przybilla et al. (2008)]. Although one cannot exclude a
possibility that BHs attain a kick at birth, we note that there is
no evidence that the kick magnitude could be as large as required by
the above considerations (see Nelemans et al. 1999; Fryer \&
Kalogera 2001; Gualandris et al. 2005).

Thus, we found that to explain the peculiar velocity of HD\,271791
the magnitude of the kick attained by the $\la 10 \, \msun$ BH
should be unrealistically large ($\geq 750-1200 \,
\kms$)\footnote{Note that the smaller the mass of the BH the larger
the kick is required to accelerate the secondary to a given
velocity.}, that makes the binary-SN ejection scenario highly
unlikely (cf. Gvaramadze 2007; Gvaramadze \& Bomans 2008).

\section{HD\,271791: dynamical ejection scenario}

The second basic mechanism responsible for the origin of runaway
stars is based on dynamical three- or four-body interactions in
dense stellar systems (Poveda et al. 1967; van Albada 1968; Aarseth
1974; Gies \& Bolton 1986). Below, we discuss three possible
channels for producing high-velocity runaways within the framework
of the dynamical ejection scenario.

The first possibility is that the high-velocity stars are the
product of breakup of unstable close triple systems (Szebehely 1979;
Anosova, Colin \& Kiseleva 1996), e.g. hierarchical triple stars
with the ratio of the semimajor axes of the outer and the inner
binaries $a_{\rm out} /a_{\rm in} \la 3-4$ (Kiseleva, Eggleton \&
Anosova 1994; Mardling \& Aarseth 2001). Such systems are
dissociated within several tens of crossing times and leave behind a
more tightly bound binary and a single (usually the least massive)
star, escaping with a velocity $V_{\rm esc} \sim (|E_0 |/m)^{1/2}$,
where $|E_0|\propto M^2 /R$ is the total energy of the system with
mass $M= m_1 + m_2 +m_3$ and scalelength $R \, (\sim a_{\rm in})$,
and $m_3 \, (< m_1 , m_2)$ is the mass of the ejected star (Valtonen
\& Mikkola 1991; Sterzik \& Durisen 1995). For a triple star
consisting of the inner binary with main-sequence components with
mass $m_1 =m_2 =50 \, \msun$ and $a_{\rm in} \simeq \, 40 \rsun$
(i.e. the binary components are close to filling their Roche lobes)
and the third star with mass $m_3=10 \, \msun$, the ejection speed
of the latter star could be $\sim 800 \, \kms$.

To reconcile this ejection scenario with the presence of
nucleosynthetic products in the atmosphere of HD\,271791, one should
assume that (i) the unstable triple star was formed due to the close
encounter between two (massive) binaries (e.g. Mikkola 1983), (ii)
HD\,271791 was a secondary component of one of these binaries, and
(iii) by the moment of the encounter, the binary containing
HD\,271791 has experienced SN explosion and remained bound [i.e. the
stellar supernova remnant (BH) received a small or no kick at
birth]. The requirement that HD\,271791 was a member of a post-SN
binary should also be fulfilled in two other dynamical processes
discussed below.

The second possibility is that the high-velocity runaways
originate through the interaction between two massive hard (Hills
1975; Heggie 1975) binaries (Mikkola 1983; Leonard \& Duncan
1990). The runaways produced in binary-binary encounters are
frequently ejected at velocities comparable to the orbital
velocities of the binary components (e.g. Leonard \& Duncan 1990)
and occasionally they can attain a velocity as high as the escape
velocity, $V_{\rm esc} =(2GM_{\ast} /R_{\ast} )^{-1/2}$, from the
surface of the most massive star in the binaries (Leonard 1991).
For the upper main-sequence stars with the mass-radius
relationship (Habets \& Heintze 1981), $R_{\ast} =0.8(M_{\ast}
/\msun )^{0.7} \rsun$, where $R_{\ast}$ and $M_{\ast}$ are the
stellar radius and the mass, one has $V_{\rm esc} \simeq 700 \,
\kms (M_{\ast}/\msun)^{0.15}$ (e.g. Gvaramadze 2007), so that the
ejection velocity could in principle be as large as $\simeq
1100-1200 \, \kms$ if the binaries contain at least one star of
mass of $20-40 \, \msun$. Numerical scattering experiments
performed by Leonard (1991) showed that about 4 per cent of
runaways produced by binary-binary interactions have velocities of
$\simeq 0.5 V_{\rm esc}$ (i.e. $\simeq 550-600 \, \kms)$, which is
enough to explain the Galactic rest-frame velocity of HD\,271791.

The third possibility is that the high-velocity runaway stars attain
their peculiar velocities in the course of close encounters between
massive hard binaries and a very massive star (Gvaramadze 2007),
formed through runaway collisions of ordinary massive stars in dense
star clusters (Portegies Zwart et al. 1999; Portegies Zwart \&
McMillan 2002; G\"{u}rkan et al. 2004). An essential condition for
the formation of very massive stars is that the runaway process
should start before the most massive stars in the cluster end their
lifetimes (e.g. G\"{u}rkan et al. 2004), i.e. the time-scale for
core collapse in the cluster (e.g. Gvaramadze et al. 2008),
\begin{eqnarray}
t_{\rm cc} \simeq 3 \, {\rm Myr} \, \left({M_{\rm cl} \over 10^4 \,
\msun }\right)^{1/2} \left({r_{\rm h} \over 1\, {\rm
pc}}\right)^{3/2} \left({\langle m \rangle \over
\msun}\right)^{-1} \nonumber \\
\times \left({\ln \Lambda \over 10}\right)^{-1} \, , \nonumber
\end{eqnarray}
where $M_{\rm cl}$ and $r_{\rm h}$ are the total mass and the
characteristic (half-mass) radius of the cluster, $\langle m
\rangle$ is the mean stellar mass and $\ln \Lambda \simeq 10$ is the
Coulomb logarithm, should be less than $\simeq 3-4$ Myr.
Observations show that the majority of star clusters are very
compact ($r_{\rm h} <1$ pc) at birth (e.g. Kroupa \& Boily 2002) so
that it is conceivable that an appreciable fraction of them evolves
through a collisional stage and form very massive stars. Simple
estimates show that our Galaxy can currently host about 100 star
clusters with a mass $M_{\rm cl} \geq 10^4 \, \msun$ (Gvaramadze et
al. 2008). All these clusters can potentially produce very massive
stars and thereby contribute to the origin of high-velocity runaway
stars.

A close encounter with the very massive star results in a tidal
breakup of the binary\footnote{Note that most of mass of very
massive stars is concentrated in a dense and compact core. According
to Yungelson et al. (2008; also Yungelson, personal communication),
the 99 per cent of mass of a $500 \, \msun$ star is confined within
a sphere of radius of $\sim 30 \, \rsun$, so that in the process of
tree-body encounter with a binary the very massive star could be
considered as a point mass (cf. Gvaramadze 2007).}, after which one
of the binary components becomes bound to the very massive star
while the second one recoils with a high velocity, given by (Hills
1988; Yu \& Tremaine 2003):
\begin{eqnarray}
V_{\infty} \simeq 500 \, \kms \, \left({M_{\rm VMS} \over 100 \, \msun}
\right)^{1/6} \left({a' \over 30\, \rsun} \right)^{-1/2} \nonumber \\
\times \left({M_1\over 10 \,\msun} \right)^{1/3}\, ,
\end{eqnarray}
where $M_{\rm VMS}$ is the mass of the very massive star and $a'$ is
the post-SN binary semimajor axis. It follows from equation\,(2)
that, to explain the peculiar velocity of HD\,271791 of $\simeq
400-600 \, \kms$, the mass of the very massive star should be $\ga
100-300 \, \msun$ [the first figure corresponds to the mass of the
most massive star formed in a `normal' way in a cluster with a mass
$M_{\rm cl} \simeq 10^4 \, \msun$ (Weidner \& Kroupa 2006)]. Note
that the weak dependence of $V_{\infty}$ on $M_{\rm VMS}$ implies
that the velocities of $\simeq 400 \, \kms$ can in principle be
produced by three-body encounters with ordinary massive stars of
mass of $\simeq 40 \, \msun$. The above estimates can be supported
by the results of three-body scattering experiments which showed
that $\ga 3$ per cent of encounters between hard massive binaries
and a very massive star of mass of $200-300 \, \msun$ produce
runaways with $V_{\infty} \geq 500-600 \, \kms$ (Gvaramadze,
Gualandris \& Portegies Zwart 2009).

Note also that the requirement that HD\,271791 was a member of a
post-SN binary does not contradict to our proposal that this star
can attain its high speed via a three-body encounter with a very
massive star (i.e. with the star more massive than the primary star
in the original binary). The merging of ordinary stars results in
effective rejuvenation of the collision product (e.g. Meurs \& van
den Heuvel 1989) so that the very massive star could still be on the
main sequence when the most massive ordinary stars start to explode
as SNe (e.g. Portegiez Zwart et al. 1999).

\section{Summary}

We have discussed the origin of the extremely high-velocity early
B-type star HD\,271791 within the framework of the binary-supernova
and the dynamical ejection scenarios. A common feature in these
competing scenarios for producing runaway stars is that HD\,271791
was a secondary of a tight massive binary and that its surface was
enriched in $\alpha$-process elements after the primary star
exploded in a supernova. Przybilla et al. (2008) favoured the
binary-supernova scenario and suggested that HD\,271791 attained its
high speed due to the disintegration of the binary caused by the
asymmetric supernova explosion. We showed, however, that to explain
the space velocity of HD\,271791 within the framework of this
scenario the kick velocity received by the stellar supernova remnant
(a $\sim 10 \, \msun$ black hole) should be extremely large, $\geq
750-1200 \, \kms$. Since there is no evidence that black holes can
attain kicks of this magnitude, we consider the binary-supernova
scenario for the origin of HD\,271791 as highly unlikely. Instead,
we proposed that the post-supernova binary remained bound and that
the high speed of HD\,271791 is due to a strong dynamical encounter
between this binary and another massive binary or a very massive
star (formed via runaway merging of ordinary stars in the core of
the parent star cluster). We argue that similar dynamical processes
could also be responsible for the origin of other hypervelocity
stars and therefore expect that the future proper motion
measurements for these objects will show that some of them were
expelled from the Galactic disc.

\section{Acknowledgements}
I am grateful to L.\,R.\,Yungelson for useful discussions and to the
anonymous referee for an advice allowing me to improve the content
of the Letter. This work was partially supported by the Deutsche
Forschungsgemeinschaft.

\end{document}